\documentclass[a4paper,fleqn]{cas-sc}

\usepackage[square,numbers]{natbib}

\usepackage{amssymb}
\usepackage[utf8]{inputenc}
\usepackage[export]{adjustbox}
\usepackage{amsfonts}
\usepackage{amsmath, amssymb}
\usepackage{bm}

\newcommand{\pdiff}[2]{\frac{\partial #1}{\partial #2}}

\newcommand{\mean}[1]{\langle #1 \rangle}

\newcommand{\rev}[1]{{\color{black}#1}}

\begin{document}

\let\WriteBookmarks\relax
\def\floatpagepagefraction{1}
\def\textpagefraction{.001}
\shorttitle{LES of a classical hydraulic jump: Influence of modelling parameters on the predictive accuracy}
\shortauthors{T. Mukha, S.K. Almeland, and R.E. Bensow}

\title [mode = title]{LES of a classical hydraulic jump: Influence of modelling parameters on the predictive accuracy }

\author[1]{Timofey Mukha}[orcid=0000-0002-2195-8408]
\ead{timofey@chalmers.se}

\credit{Conceptualization of this study, Methodology, Investigation, Visualization, Data curation, Writing - Original draft preparation}

\address[1]{Chalmers University of Technology, Department of Mechanics and Maritime Sciences, Hörsalsvbägen 7A, SE-412 96 Gothenburg, Sweden}

\author[2]{Silje Kreken Almeland}[orcid=0000-0003-3106-0623]
\ead{silje.k.almeland@ntnu.no}
\address[2]{Norwegian University of Science and Technology, Department of Civil and Environmental Engineering, NO-7491 Trondheim, Norway}
\cormark[1]
\credit{Conceptualization of this study, Methodology, Investigation,  Resources, Writing - review \& editing}

\author[1]{Rickard E. Bensow}[orcid=0000-0002-8208-0619]
\ead{rickard.bensow@chalmers.se}

\credit{Conceptualization of this study, Funding acquisition, Project administration, Resources, Supervision, Writing - review \& editing}

\cortext[cor1]{Corresponding author}

\begin{abstract}
Results from large-eddy simulations of a classical hydraulic jump at inlet Froude number 2 are reported.
The computations are performed using the general-purpose finite-volume based code OpenFOAM\textsuperscript{\textregistered}, and the primary goal is to evaluate the influence of modelling parameters on the predictive accuracy, as well as establish associated best-practice guidelines.
A benchmark simulation on a dense computational mesh is conducted, and good agreement with existing reference data is found.
The remaining simulations cover different selections of modelling parameters: geometric vs algebraic interface capturing, three mesh resolution levels, four choices of the convective flux interpolation scheme.
Geometric interface capturing leads to better accuracy but deteriorated numerical stability and increased simulation times.
Interestingly, numerical dissipation is shown to systematically improve the results, \rev{both in terms of accuracy and stability}.
The \rev{densest} of the three grids, which is twice as coarse as the grid used in the benchmark simulation, was found to be sufficient for faithfully reproducing all the considered quantities of interest.
The recommendation is therefore to use this grid, geometric interface capturing, and a second-order upwind scheme for the convective fluxes.

\end{abstract}


\begin{keywords}
	Hydraulic jump \sep
	Large-eddy simulation \sep
	CFD \sep
	OpenFOAM
\end{keywords}

\maketitle
\section{Introduction} \label{sec:intro}

A hydraulic jump is an abrupt change in the water depth accompanying the transition of the flow in a shallow canal from super- to subcritical.
This transition causes energy dissipation, which defines the application of hydraulic jumps in engineering.
In fact, according to~\cite{Bayon-Barrachina2015}, hydraulic jumps are the most commonly used energy dissipator in hydraulic structures.
This motivates the significant attention this class of flows received from the scientific community.
Hydraulic jumps have been the subject of a multitude of studies, both experimental and numerical, a recent review of which can be found in~\cite{Valero2018, Viti2018}.
Most works focus on the so-called `classical' hydraulic jump (CHJ), which occurs in a smooth horizontal rectangular channel.

\begin{figure}[htp!]
	\includegraphics[width=\linewidth]{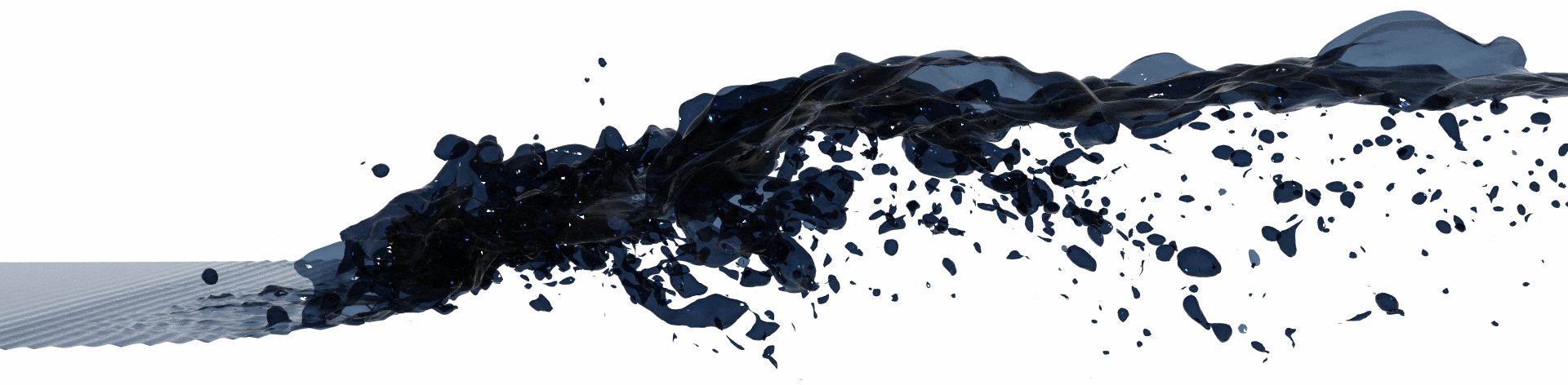}
	\caption{Classical hydraulic jump at $Fr_1 = 2$, a snapshot of the air-water interface.}
	\label{fig:intro}
\end{figure}

An illustration of the air-water interface in a CHJ is shown in Figure~\ref{fig:intro}.
The topology of the interface is complex and rapidly evolving, which can be fully appreciated by looking at the animations found in the supplementary material to this article.
The interface dynamics are driven by a recirculating motion---the so called roller--- which leads to overturning waves occurring across the jump.
Consequently, a significant amount of air is entrained.
A detailed discussion of the entrainment mechanism can be found in~\cite{Mortazavi2016a}.
The flow in the jump is also highly turbulent, with a turbulent shear layer forming below the roller and interacting with it.

The main physical parameter of the CHJ is the inlet Froude number, $\text{Fr}_1$, computed based on the water inlet velocity, $U_1$, and its depth, $d_1$.
Several classifications of the jump's behaviour based on $\text{Fr}_1$ can be found in the literature, see~\cite{Valero2018} and the references therein.
The most stable CHJs occur when $\text{Fr}_1 \in [4, 9]$.
The rate of air entrainment also depends on the Froude number, and at higher $\text{Fr}_1$ the level of aeration is increased.
In spite of the flow's complexity, given the parameters of the inflow, some of its properties can be easily derived analytically based on control volume analysis, see e.g.~\cite[p. 250]{Kundu2008}.
This includes the water depth after the jump, $d_2=0.5d_1\left((1 + 8\text{Fr}_1^2)^{0.5} - 1\right)$.

From a numerical perspective, the CHJ represents an extremely challenging test of predictive capabilities for multiphase modelling approaches.
A suitable model should be able to capture fast and complex topology changes taking place across a wide range of spatial scales.
Accurate turbulence modelling is also necessary and, in particular, the possibility to properly account for its interaction with the multiphase structures.
On the other hand, the geometric simplicity of the case makes mesh generation easy, and the abundance of published experimental data makes validation easier.
Furthermore, data from direct numerical simulation (DNS)~\cite{Mortazavi2016a} is also available.

A compilation of previous numerical studies of the CHJ, classified by turbulence modelling approach, and also~$\text{Fr}_1$, can be found in~\cite{Viti2018}.
The majority of works are based on two-equation Reynolds-averaged Navier-Stokes (RANS) turbulence models and the Volume of Fluid (VoF) method for capturing the interface.
Note that in RANS it is assumed that there is a clear scale separation between the modelled turbulent motion and other types of unsteadiness.
In the case of a CHJ, this is unlikely to take place due to the direct interaction between turbulence and multiphase structures, e.g. entrained bubbles.
Generally, it is unclear whether the topological changes in the flow occur significantly slower than the integral time scales of turbulence.
A possibility for resolving this inconsistency is keeping the resolution coarse enough for the interface to remain steady, and introduce an explicit model for air entrainment, as done in~\cite{Ma2011}.
However, these theoretical difficulties do not imply that RANS cannot be used to obtain useful results.
On the contrary, as summarized in~\cite{Viti2018}, RANS is capable of predicting $d_2$, the mean location of the interface, and the length of the roller with $<5\%$ relative error.

In order to get new physical insights, and get an accurate picture of the turbulent motion inside the jump, scale-resolving turbulence modelling approaches can be used.
Only a few studies report results from such simulations.
The DNS by Mortazavi et al.~\cite{Mortazavi2016a} has already been mentioned above, and represents an important milestone.
In~\cite{Jesudhas2018}, Detached-Eddy Simulation (DES) was used.
A detailed analysis of the flow is given, in particular, a quadrant decomposition of the turbulent shear stress is considered, as well as high-order statistical moments of the velocity field.
A Large-Eddy Simulation (LES) of a CHJ was conducted as part of the study by Gonzalez and Bombardelli, which also includes RANS simulations~\cite{Gonzalez2005}.
Unfortunately, due to the reference being a short conference abstract, results are only discussed superficially.
In~\cite{Lubin2009} the authors report on an unsuccessful attempt to conduct a LES: the location of the jump could not be stabilized.
Finally, in~\cite{Ma2011}, which was already mentioned in the context of RANS, results from DES modelling are also discussed.
However, the simulation in question is not a DES in the classical sense, i.e.~not a fully resolved LES outside the RANS region.
Instead, a rather coarse mesh is used and an air entrainment model is employed.
Nevertheless, this DES yielded more accurate results than the corresponding RANS.

In summary, only two articles~\cite{Mortazavi2016a, Jesudhas2018} contain detailed reports on scale-resolving simulations of the CHJ to date.
Nevertheless, with the increase of available computing power, it can be expected that LES and its hybrids will find wider adoption in hydraulic engineering in the near future.
This transition is already well underway in, for example, the automotive and aerospace industries.
From a practical perspective, an important step is to establish guidelines for the selection of the most important LES modelling parameters.
Of immediate interest is to form them for the particular case of solvers based on finite-volume discretisation, since these are currently the workhorse of industrial computational fluid dynamics.
In this numerical framework, the two arguably most important LES input parameters are the density of the grid and the numerical scheme used for interpolating convective fluxes.
Both control the capability of the LES to resolve turbulent structures, and also the stability of the simulation.
In the case of multiphase flow, the choice of the interface capturing scheme is also important.
The goal of this paper is to quantify the effects of these three parameters on the various quantities of interest.
To that end, results from an LES campaign, consisting of 25 simulations of a CHJ at $\text{Fr}_1 = 2$, are presented.
The campaign covers four different mesh resolution levels, and two VoF approaches: algebraic and geometric.
The diffusivity of the convective flux interpolation scheme is also controlled, and four diffusivity levels are considered.
All the simulation results, including ready-to-run simulation cases, are made available as a supplementary dataset.\footnote{DOI: \texttt{10.6084/m9.figshare.12593480}}

The remainder of the article is structured as follows.
Section~\ref{sec:cfd} presents the computational fluid dynamics methods used in the paper.
The setup of the CHJ simulations is presented in Section~\ref{sec:setup}.
The results of the simulation campaign are shown and analysed in Section~\ref{sec:experiments}.
Finally, conluding remarks are given in Section~\ref{sec:conclusions}.

\section{Computational fluid dynamics methods} \label{sec:cfd}

\subsection{Governing equations}
The Volume of Fluid (VoF) method~\cite{Hirt1981} is used to simulate the flow.
Accordingly, a single set of conservation equations is solved for both fluids, and the phase is distinguished based on the values of the volume fraction of the liquid, $\alpha$.
The momentum and continuity equations read as follows,
\begin{align}
\pdiff{\rho u_i}{t}+\pdiff{}{x_j}\left(\rho u_iu_j \right) & = -\pdiff{p_{\rho gh}}{x_i} - g_ix_i\pdiff{\rho}{x_i} +  \pdiff{}{x_j}\left( \mu \left( \pdiff{u_i}{x_j} + \pdiff{u_j}{x_i} \right) \right) + f_s,\quad (i=1,2,3) \label{eq:lesmom} \\
\label{eq:lescont}
\pdiff{u_j}{x_j} & =  0.
\end{align}
Here, summation is implied for repeated indices, $u_i$ is the velocity, $\rho$ is the density, $\mu$ is the dynamic viscosity,  $g_i$ is the standard acceleration due to gravity, $p_{\rho gh} = p - \rho g_ix_i$ is the dynamic pressure, and $f_s$ is the surface tension force.
The latter is accounted for using the Continuous Force Model~\cite{Brackbill1992}:
\begin{equation}
\label{eq:surface_tension}
f^s_i = \sigma \kappa \pdiff{\alpha}{x_i}.
\end{equation}
Here, $\sigma$ is the surface tension coefficient and $\kappa = \partial n^f_i/\partial x_i$ is the curvature of the interface between the two phases, where $n^f_i$ is the interface unit-normal, which is computed as follows,
\begin{equation} \label{eq:normal}
n^f_i= \pdiff{\alpha}{x_i} \bigg/ \left( \bigg\vert \pdiff{\alpha}{x_i}  \bigg\vert + \delta_N \right) .
\end{equation}
Here, $\delta_N$ is a small number added for the sake of numerical stability.

Equations~\eqref{eq:lesmom}-\eqref{eq:lescont} must be complemented with an interface capturing approach in order to compute the distribution of $\alpha$.
Methodologies for this are discussed in the next subsection.
Given the values of $\alpha$, the local material properties of the fluid are computed as
\begin{align}
& \rho = \alpha\rho_{1} + (1 - \alpha)\rho_{2},
& \mu = \alpha\mu_{1} + (1 - \alpha)\mu_{2},
\end{align}
where the indices $1$ and $2$ are used to refer to the liquid and gas properties, respectively.

\subsection{Interface capturing methods}
As discussed above, it is necessary to introduce a method for computing the evolution of $\alpha$, i.e.~capture the location of the interface between the two fluids.
Here, two different approaches to this are considered.
The first is algebraic, meaning that a transport equation for $\alpha$ is solved:
\begin{equation}
\label{eq:alpha_mules}
	\pdiff{\alpha}{t} + \pdiff{u_j \alpha }{x_j} + \pdiff{}{x_j}\left(u^r_j (1 -\alpha)\alpha\right) = 0.
\end{equation}
The last term in the equation is artificial and its purpose is to introduce additional compression of the interface.
To that end, the direction of $u_i^r$ is aligned with the interface normal, $n^f_i$.
The magnitude of $u^r_i$ is defined as $C_\alpha |u_i|$, where $C_\alpha = 1$ is an adjustable constant.

Special treatment of the convective term in~\eqref{eq:alpha_mules} is necessary in order to ensure that $\alpha$ is bound to values between 0 and 1.
Typically, a total-variation diminishing (TVD) scheme is chosen to compute the convective flux, but this can be insufficient because the TVD property of such schemes is, in fact, only strictly valid for one-dimensional problems.
An additional flux limiting technique, referred to as MULES, is used to rectify this.
While we omit discussing MULES in detail and instead refer the reader to~\cite{Damian2013}, we note that it is based on the idea of Flux Corrected Transport and the work of Zalesak~\cite{Zalesak1979}.
It should also be mentioned that two variations of MULES are available in OpenFOAM\textsuperscript{\textregistered}, explicit and semi-implicit.
Using the latter sometimes allows to keep the simulation stable for CFL numbers larger than one.

The second approach belongs to the class of geometric VoF methods, and is referred to as isoAdvector.
The details on isoAdvector can be found in~\cite{Roenby2016}, here we provide a brief summary of the key steps of the algorithm.
In contrast to algebraic VoF, here the surface of the interface is explicitly reconstructed at each time-step.
Within each cell, it is represented by a plane, and the reconstruction algorithm ensures that it divides the cell volume consistently with the local value of $\alpha$.
To predict the location of the interface at the next time-step, it is advected along the direction of the interface normal.
For each cell, the advection velocity is obtained using linear interpolation from the vertices of the cell onto the centroid of the interface-plane.
Then, based on the predicted new location of the interface, the change in $\alpha$ is computed.

Comparing the two approaches, one can generally say that geometric VoF can be expected to be more accurate, yet more computationally demanding.
Quantifying these differences for the case of the hydraulic jump is one of the goals of the present paper.
A significant drawback of the algebraic VoF is the necessity to choose the convection scheme for $\alpha$, which can have a large influence on the results.
Selecting the values of model constants, such as $C_\alpha$, also represents a difficulty.

\subsection{Numerical methods} \label{sec:numerics}

The computations are performed using the open-source CFD software OpenFOAM\textsuperscript{\textregistered} version 1806.
This code is based on cell-centred finite-volume discretization, which can currently be considered standard for industrial CFD.
Two solvers distributed with OpenFOAM\textsuperscript{\textregistered} were employed, corresponding to the two VoF methodologies discussed above.
For algebraic VoF, the solver \texttt{interFoam} was used, whereas the isoAdvector is implemented in the \texttt{interIsoFoam} solver.
Here we omit the particulars regarding the solver algorithms, but note that they are based on the PISO~\cite{Issa1986} pressure-velocity coupling procedure.
For a detailed discussion we refer the reader to the following thesis works~\cite{Rusche2002, Damian2013}.

A key component of the finite-volume method are the spatial interpolation and time integration schemes.
For spatial interpolation, the goal is to obtain the values of the unknowns at the cell face centroids based on the values at the centroids of the cells.
The most trivial choice is using linear interpolation, which is second-order accurate.
This scheme can be applied to interpolation of diffusive fluxes without negative side-effects.
Unfortunately, when applied to convective fluxes, linear interpolation leads to a dispersive error.
In spite of this, in single-phase LES and DNS it is common practice to use this scheme anyway because the high density of the mesh, in combination with a small time-step, allows to avoid any significant contamination of the solution.
On the other hand, in industrial flow simulations it is quite common to use a second-order upwind scheme.
Although also unbounded, the error introduced by this scheme is dominated by a dissipative term, which facilitates the stability of the simulation but negatively affects the capability to resolve small-scale turbulent motions.
In this work, a linear blending of these two interpolation schemes will be considered.
The following weights for the linear upwind scheme will be tested: 10\%, 25\%, 50\%, and 100\%.
For simplicity, this weight will be referred to as `the amount of upwinding' in the remainder of the paper.

For time integration, both solvers have the option of using a first-order implicit Euler scheme.
In \texttt{interFoam}, the Crank-Nicholson scheme can also be used, as well as a linear blending of Crank-Nicholson and Euler.
In \texttt{interIsoFoam}, one can instead use a second-order accurate backward-differencing scheme.
Unfortunately, it was only possible to keep the simulations stable using the Euler scheme.
However, since the employed time-step sizes are kept low, it is anticipated that the numerical errors are dominated by the spatial interpolation errors, whereas the time-integration error plays a smaller role.

Finally, in the case of MULES, a scheme has to be chosen for the convection of $\alpha$.
Here a TVD scheme using the van Leer limiter is selected to that end, see~\cite[p. 170]{Versteeg2007} for the definition.
The selected limiter results in a more diffusive scheme than some alternatives, but here the artificial compression term in~\eqref{eq:alpha_mules} remedies that.

\subsection{Turbulence modelling} \label{sec:turbulence}

In order to obtain the governing equations for LES based on the employed two-phase flow model, spatial filtering should be formally applied to equations~\eqref{eq:lesmom}-\eqref{eq:lescont}, as well as~\eqref{eq:alpha_mules} in the case of algebraic VoF.
Following standard practice, we use implicitly-filtered LES, letting the finite volume grid act as the spatial filter.
The associated filter size is equal to the cubic root of the local computational cell volume.

Filtering leads to the appearance of the subgrid stress (SGS) term in the momentum equation~\eqref{eq:lesmom}.
For several reasons, here we choose to ignore this term instead of modelling it.
One reason is the relatively dissipative numerical schemes employed in the solution procedure.
When a high amount of upwinding is present, it can be expected that the numerical dissipation is comparable in magnitude to that produced by a Boussinesq-type SGS model.
This stacking of artificial and modelled dissipation leads to deterioration of accuracy.
Another reason is the overall negative experience of the authors with the SGS models implemented in OpenFOAM\textsuperscript{\textregistered}.
Some results for turbulent channel flow and several available models can be found in~\cite{Mukha2016}.
In that study, only the dynamic $k$-equation model~\cite{Kim1995} does not lead to worsened results compared to not using a model at all.
Also, one of the goals of this paper is to see what accuracy can be achieved on coarse grids, which are technically not suitable for LES but may nevertheless appear in an industrial setting due to limitations in computing resources.
Using an SGS model that is designed to model a specific part of turbulent spectrum is unlikely to be fruitful in this context.
\rev{Finally, it should be noted that the existing closures where designed for single-phase flows,	and do not account for the interaction effects between multiphase and turbulent structures.
Detailed investigations of the impact of this on the predictive accuracy have not yet been reported in the literature.
Consequently, using such SGS models in hydraulic jump simulations can be called into question.}

\subsection{Instability sources in VoF simulations} \label{sec:instable}
Compared to single-phase LES, numerical stability in LES-VoF simulations can be significantly harder to achieve.
As discussed in Section~\ref{sec:experiments}, for certain combinations of the grid resolution, VoF methodology, and amount of upwinding, the CHJ simulations diverged.
It is therefore appropriate to briefly review the main additional sources of numerical instability intrinsic to the considered multiphase modelling approach.

The Continuous Force Model, see \eqref{eq:surface_tension}-\eqref{eq:normal}, used for the surface tension force computation can lead to parasitic currents across the interface between the phases.
An illustration of such currents produced in an \texttt{interFoam} simulation of a single rising bubble can be found in~\cite{Cano-Lozano2015}.
The source of the currents is the numerical imbalance between the pressure gradient across the interface and the surface tension.
When the velocity of the parasitic current becomes large, the simulation may crash.
A multitude of improvements to the Continuous Force Model have been proposed, ranging from more accurate curvature estimation algorithms to more robust discrete handling of the balance between pressure and surface tension forces, see, for example,~\cite{Popinet2009}.
Unfortunately, none of these have been implemented and publicly released in OpenFOAM\textsuperscript{\textregistered}, although there are ongoing efforts~\cite{Scheufler2019}.
Interestingly, in~\cite{Scheufler2019} the author mentions that the sharper interface obtained using \texttt{isoAdvector} actually increases the magnitude of the parasitic currents.

The second source of instabilities is the treatment of the gravity term in the momentum equation~\eqref{eq:lesmom}.
When a segregated pressure-velocity coupling algorithm, such as PISO, is used, a numerical imbalance between the dynamic pressure gradient and the density gradient terms can occur, which will be compensated by an acceleration of the fluid~\cite{Vukcevic2017}.
This can lead to a strong increase of velocity magnitude in the gas above the interface, due to its low density.
For this reason, it is not uncommon to artificially increase the density of the gas to facilitate stability.

The crucial practical consequence of the above is that one does not necessarily get a more stable simulation by refining the grid and using a more accurate interface capturing approach.
This is very different from single-phase incompressible LES, in which dampening numerical instabilities by using a denser grid or a smaller time-step is a common strategy.
It should also be mentioned that it is not possible to predict when the discussed instabilities will take place.
Some of the CHJ simulations conducted as part of this study were well under way when the destabilizing velocity overshoots occurred, leading to loss of tens of thousands of core-hours worth of computing time.
An even more unfortunate scenario is when a very strong spurious current takes place, but no crash occurs.
The simulation finishes, but the results are unpublishable because the computed statistical moments of velocity are contaminated.
In our simulations, the solver would sometimes exhibit surprising resilience and survive currents that are 3 orders of magnitude stronger than the characteristic velocity scale of the flow.
It is therefore recommended to closely monitor the maximum velocity values in the course of the simulation.

\section{Simulation setup} \label{sec:setup}
\begin{figure}[htp!]
	\centering
	\includegraphics[width=\linewidth]{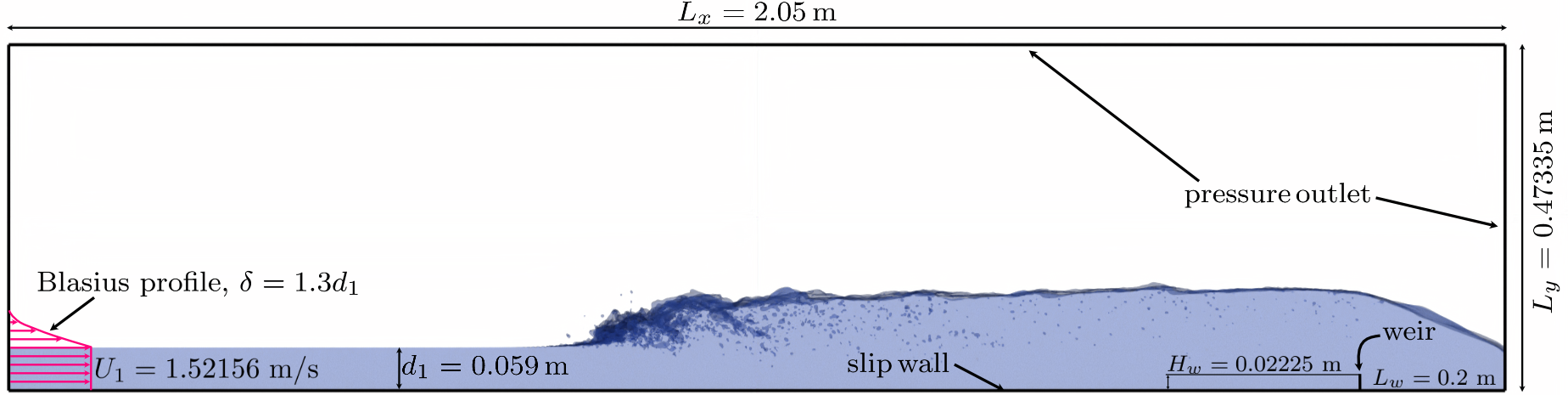}
	\caption{Simulation setup.}
	\label{fig:setup}
\end{figure}

The setup of the simulation is similar to that used in the DNS by Mortazavi et al.~\cite{Mortazavi2016a}, which is later used as reference.
An overview of the computational domain, as well as boundary conditions, can be found in Figure~\ref{fig:setup}, and Table~\ref{tab:setup} contains a full list of the setup parameters.
The main difficulty in setting up a hydraulic jump simulation is obtaining a stable jump positioned sufficiently far away from the inlet and outlet boundaries of the domain.
A common approach to facilitating the formation of the jump is by introducing a vertical barrier---a weir---some distance upstream of the outlet, see e.g.~\cite{Witt2015, Jesudhas2018,  Witt2018}.
In other works~\cite{Mortazavi2016a, Bayon2016, Bayon-Barrachina2015}, including the reference DNS, the jump is controlled by the boundary condition at the outlet of the domain.
The particular condition enforced varies among the studies.

Here, the weir approach is employed due to its simplicity.
This choice also facilitates reproducibility by making the simulation setup easier to reproduce in any CFD code, without the need to program a new boundary condition.
Test simulations were necessary to find a configuration of the domain length $L_x$, weir height $H_w$, and the streamwise position of the weir, $L_w$, in order to get a stable jump positioned roughly in the middle of the domain.
The streawmise dimension of the weir is always set to equal the size of a computational cell, which is defined below.
A simple pressure outlet is used on the downstream boundary.

At the inlet, the depth of the water, $d_1$ and the inlet velocity $U_1$, are set to enforce $\text{Fr}_1 =U_1/\sqrt{g d_1} = 2$.
In the air phase, a Blasius boundary layer profile subtracted from $U_1$ is enforced.
The thickness of the boundary layer is $\delta = 1.3 d_1$.
This matches the condition in the reference DNS.

The condition at the bottom surface is also matched and is set to a slip wall.
This allows to not spend computational resources on the boundary layer, which would have been formed if a no-slip condition were to be imposed.
In the DNS, a slip condition is also used for the top boundary.
However, we find this choice difficult to justify and instead impose a pressure outlet, mimicking the atmosphere.
Accordingly, the height of the domain is made significantly larger than in the DNS as well.
Some test simulations with a slip applied to the top boundary were nevertheless conducted, and the changes in the obtained solution were not significant.

The spanwise extent of the domain, $L_z$, was set to match the DNS, $L_z = 4.2d_1$.
However, analysis of two-point autocorrelations of the velocity field  at selected locations (presented below) revealed that a larger $L_z$ should preferably be used.
Due to limitations in computational resources, it was not possible to extend $L_z$ in all the conducted simulations.
However, in the simulations on coarser meshes (defined below), $L_z = 8.4d_1$ was used.
One the one hand, this can be seen as an impediment to consistent evaluation of the predictive accuracy across several mesh densities.
On the other hand, simulations on coarse meshes are more prone to deteriorating in accuracy due to an insufficiently wide domain, because a coarse mesh tends to introduce spurious spatial correlations.
The latter consideration was judged to outweigh the former.

It remains to define the material properties of the fluids: their densities, kinematic viscosities, and also the surface tension coefficient.
These are adjusted to exactly match the dimensionless parameters of the DNS, which includes the Weber number, $\text{We} = \rho_1 U_1^2 d_1/\sigma = 1820$, the Reynolds number, $\text{Re} = U_1 d_1/\nu_1 = 11000$, density ratio, $\rho_1/\rho_2 = 831$, and dynamic viscosity ratio, $\mu_1/\mu_2 = 50.5$.
The corresponding dimensional values can be found in Table~\ref{tab:setup}.

\begin{table}[htp]
	\caption{Simulation parameters.}
	\label{tab:setup}
	\begin{tabular}{ll}
		\textbf{Property}                                   & \textbf{Value}                                      \\ \hline
		Water inlet height, $d_1$                           & 0.0059 m                                            \\
		Domain length, $L_x$                                & 2.05 m, $\approx 34.75 d_1$                         \\
		Domain height, $L_y$                                & 0.47355 m, $\approx 8 d_1$                          \\
		Domain width, $L_z$                                 & 0.2478 m / 0.4956 m, $4.2d_1$  / $8.4 d_1$          \\
		Weir height, $H_w$                                  & 0.02225 m                                           \\
		Distance from the weir to the outlet, $L_w$         & 0.2 m                                               \\
		Liquid density, $\rho_1$                            & 1.20012 kg/$\mathrm{m^3}$                           \\
		Gas density, $\rho_2$                               & 997.3 kg/$\mathrm{m^3}$                             \\
		Density ratio, $\rho_1/\rho_2$                      & 831                                                 \\
		Liquid kinematic viscosity, $\nu_1$                 & $8.1611213 \cdot 10^{-6}$ $\mathrm{m^2}/\mathrm{s}$ \\
		Gas kinematic viscosity, $\nu_2$                    & $1.34295 \cdot 10^{-4}$ $\mathrm{m^2}/\mathrm{s}$   \\
		Kinematic viscosity ratio, $\nu_1/\nu_2$            & 0.06077                                             \\
		Dynamic viscosity ratio, $\mu_1/\mu_2$              & 50.5                                                \\
		Surface tension coefficient, $\sigma$               & 0.07484925                                          \\
		Water inlet velocity, $U_1$                         & 1.52156 m/s                                         \\
		Inlet Froude number, $\text{Fr}_1$      			& 2                                                   \\
		Weber number, $\text{We}$						    & 1820                                                \\
		Reynolds number, $\text{Re}$      				    & $11\,000$ 									      \\ \hline
	\end{tabular}
\end{table}

Several computational meshes, varying in their density, are employed in the study.
All the meshes are fully defined in the next section, and here the general topology, which all the meshes share, is presented.
The region occupied by the jump is meshed using cubic cells.
This can be considered optimal in terms of the performance of the employed numerical algorithms.
A rapid coarsening towards the top boundary is introduced slightly above the half-height of the domain.
Similarly, the mesh is coarsened towards the outlet past the location of the weir.
Coarsening towards the inlet is also present, starting about half-way from the position of the jump to the inlet.

\section{Numerical experiments} \label{sec:experiments}

In this section, the results of the simulations are presented and discussed.
First, an overview of the simulation campaign is given in Section~\ref{sec:campaign}.
This is followed by a presentation of results from the simulation on the densest mesh and their comparison with reference DNS data in Section~\ref{sec:benchmark}.
Finally, in Section~\ref{sec:parameters}, the effects of various modelling parameters are quantified.

\subsection{Simulation campaign overview} \label{sec:campaign}

The simulation campaign consists of 25 cases, which differ in the amount of upwinding introduced by the convective flux interpolation scheme, the density of the grid, and the VoF methodology employed.

A single simulation, referred to as the benchmark, has been run on a grid with the edge of the cubic cells $\Delta x$ set to $1$ mm, which is approximately equal to the resolution used in the DNS.
With this grid, the theoretical height of the jump, $d_2 - d_1$, is discretized by 81 cells.
The size of the grid is $\approx 83$ million cells.
The algebraic VoF was used, and only 2\% upwinding was employed.
We note that the initial plan was to use the geometric VoF for the benchmark simulation due to its superior accuracy.
However, stabilizing the simulation proved difficult.
Several costly attempts were made, with the amount of upwinding gradually increased, but even with 20\% upwinding instabilities occurred.

The rest of the simulations cover the following choices for the grid resolution, $\Delta x \in [2, 3, 4]$ mm,  and amount of upwinding, $[10\%, 25\%, 50\%, 100\%]$.
We will from here on refer to the four grids used in the study as [$\Delta x1$, $\Delta x2$, $\Delta x3$, $\Delta x4$], and denote the amount of upwinding as [$u10\%$, $u25\%$, $u50\%$, $u100\%$].
For each configuration, algebraic and geometric VoF are considered, which will be referred to by the name of the key underlying algorithm, MULES and isoAdvector, respectively.
As mentioned in Section~\ref{sec:setup}, for simulations on grids $\Delta x3$ and $\Delta x4$ the value of $L_z$ was doubled.

All simulations were first run for 1 s of simulation time, after which time-averaging was commenced and continued for 11 s.
This corresponds to $\approx 283 d_1/U_1$ time-units.
By comparison, the reference DNS was averaged across 120 time-units.
To obtain the final statistical results spatial averaging along the spanwise direction was performed.
The time- and spanwise-averaged quantities will be denoted with angular brackets below, $\langle \cdot \rangle$.
Adaptive time-stepping based on the maximum value of the CFL number currently registered in the domain was used.
For simulations using MULES, the maximum CFL allowed was 0.75, whereas 0.5 was used with isoAdvector.

Further notation used in the remainder of the paper is now introduced.
The mean location of the interface is denoted as $\mean{\alpha_{0.5}}$, corresponding to the 0.5 isoline in the mean volume fraction field.
The triple $u$, $v$, $w$ is used to denote the three Cartesian components of velocity.
The location of the toe of the jump, $x_{toe}$, is defined as the streawise location at which the vertical position of $\mean{\alpha_{0.5}}$ is $1.1d_1$.
The same definition is used in the reference DNS data~\cite{Mortazavi2016a}.
The following rescaling of the coordinate system will be used: $x' = (x - x_{toe})/d_1$, $y' = y/d_1$.

\subsection{Benchmark simulation} \label{sec:benchmark}
\rev{Here the results of the benchmark simulation are compared to the DNS of Mortazavi et al.~\cite{Mortazavi2016a}.
The grid resolution in the two simulations is similar, but the setup doesn't match exactly, as pointed out in Section~\ref{sec:setup}.
There are also certain differences in the definitions of the considered quantities of interest, as discussed below.
Additionally, for certain quantities the DNS is clearly poorly converged.
Nevertheless, a qualitative and, in most cases, quantitative comparison of the results is possible, with the DNS generally considered as reference, since it was performed using more accurate numerics.}
The primary goal here is not to obtain perfect agreement, but rather to answer the principle question of whether the employed physical and numerical modelling frameworks are capable of capturing the properties of such a complicated flow.

\begin{figure}[htp!]
	\includegraphics{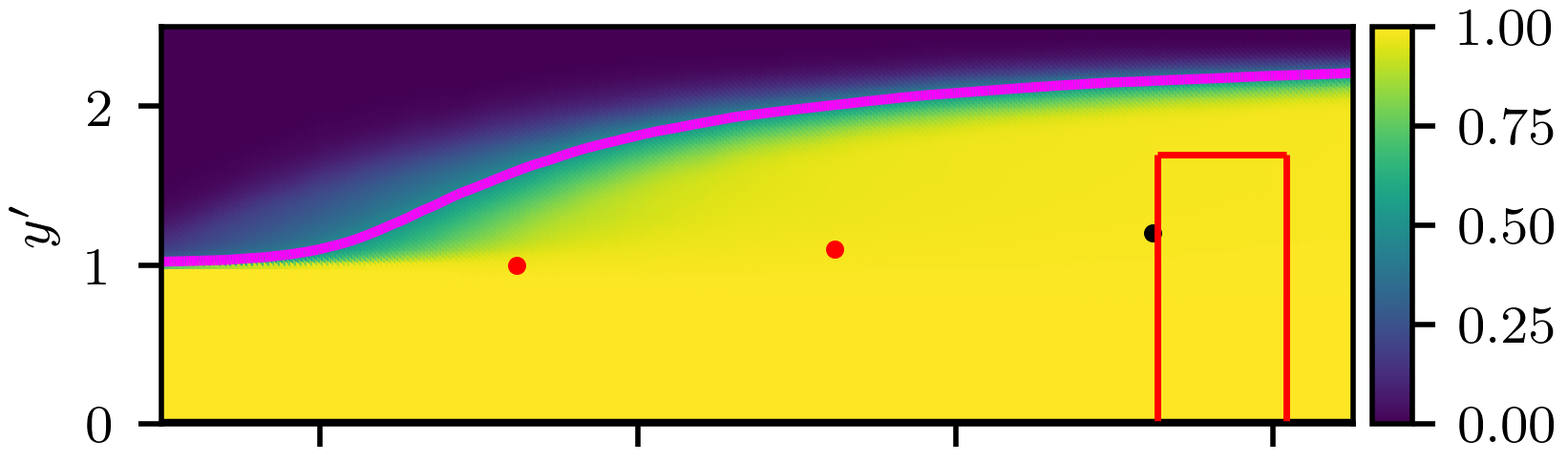}
	\includegraphics{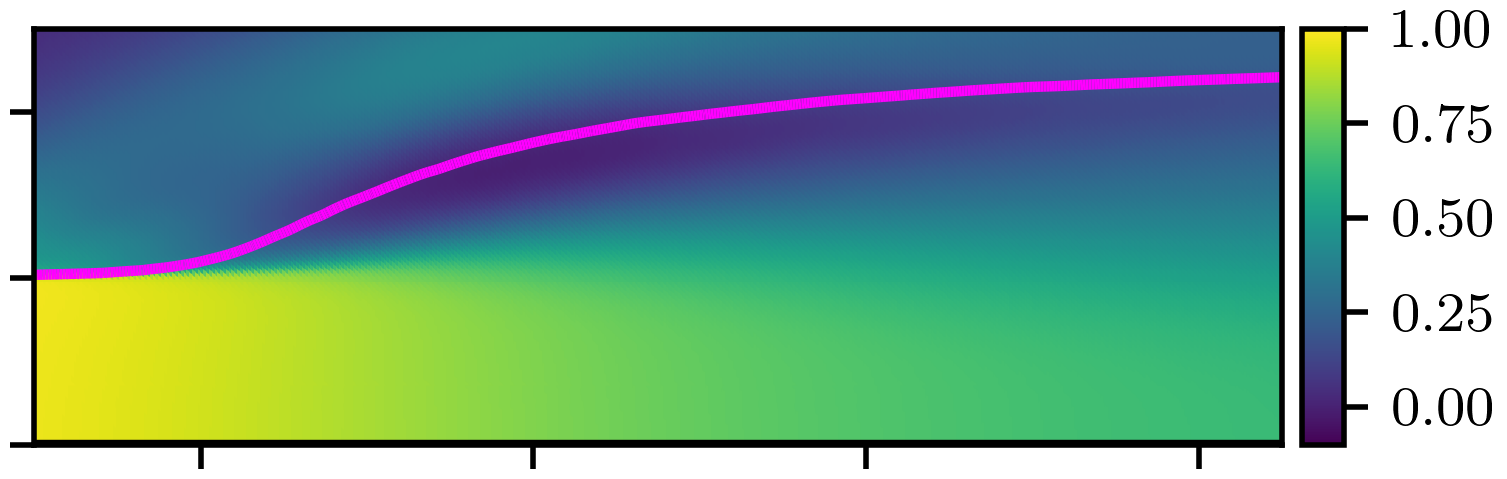}

	\includegraphics{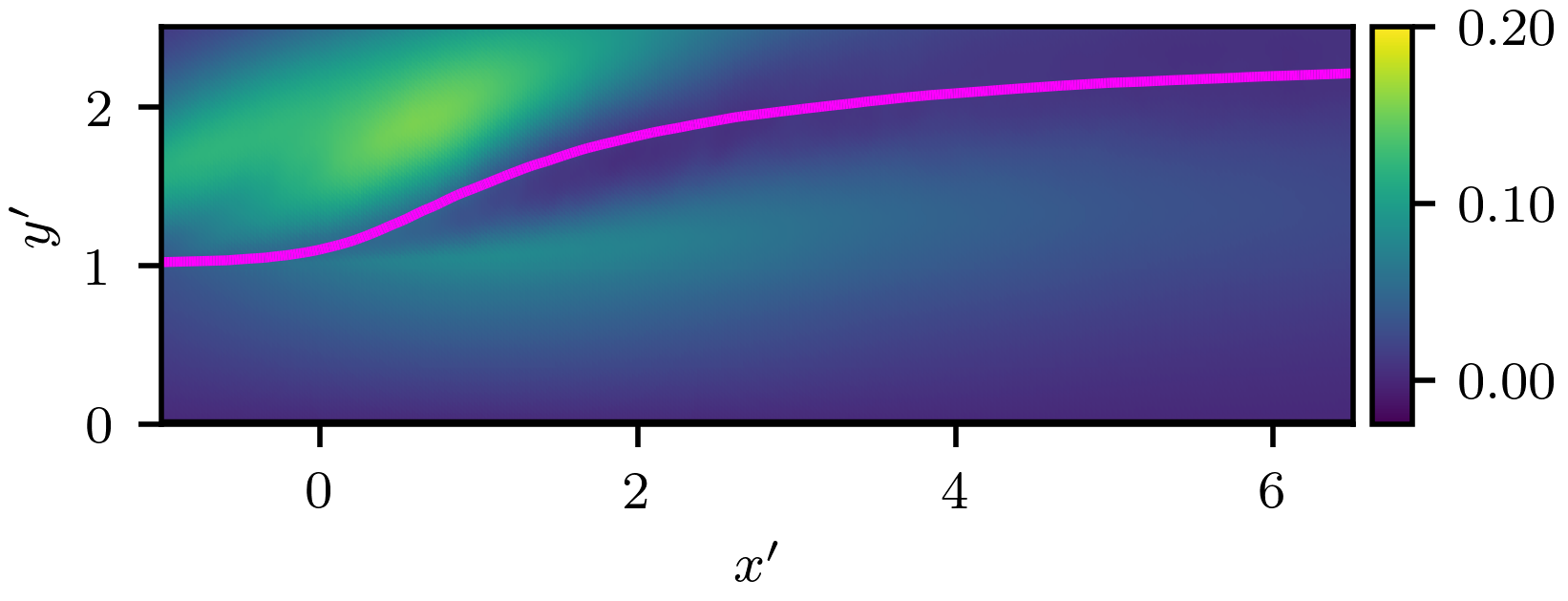}
	\includegraphics{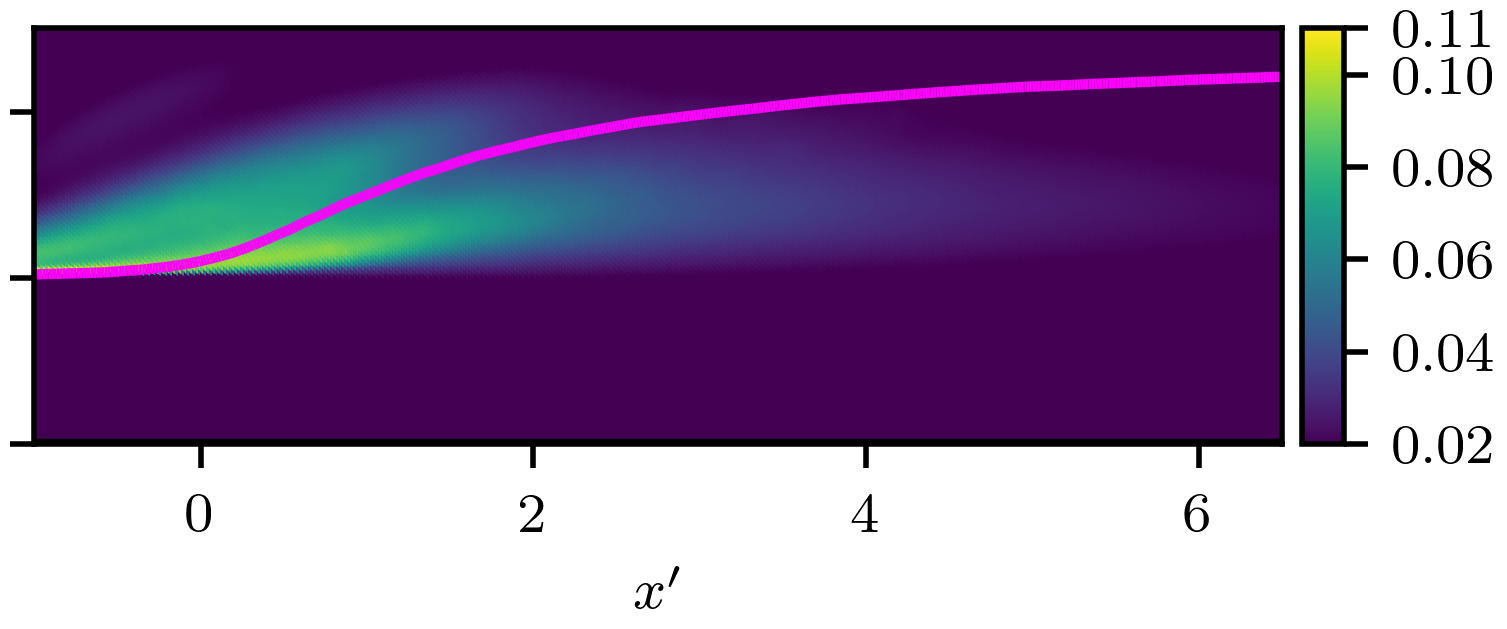}
	\caption{Distribution of $\mean{\alpha}$ (top-left), $\mean{u}/U_1$ (top-right), $\mean{v}/U_1$ (bottom left), $\mean{k}/U^2_1$ (bottom-right) in the benchmark simulation. The magenta line shows $\mean{\alpha_{0.5}}$.}
	\label{fig:bench_alpha_2d}
\end{figure}

An overview of the distribution of the main flow quantities is given first, see Figure~\ref{fig:bench_alpha_2d}.
The top-left plot shows the distribution of $\mean{\alpha}$, with the magenta line showing $\mean{\alpha_{0.5}}$.
Close to the toe of the jump, and some distance downstream, the values of $\mean{\alpha}$ are significantly lower than 1, indicating air entrainment.
The mean streamwise and vertical velocities are shown in the top-right and bottom-left plots, respectively.
As expected, the streamwise velocity is significantly lower downstream of the jump.
It is also visible how the boundary layer in the gas follows the interface, leading to an increase in the vertical velocity in a region above the toe of the jump.
Finally, the mean turbulent kinetic energy per unit mass, $\mean{k}$, is shown in the bottom-right plot.
High values are observed in the region close to the toe, with the peak directly downstream of it.
This reflects the coupling between turbulence and the air entrainment.

\begin{figure}[htp!]
	\centering
	\includegraphics{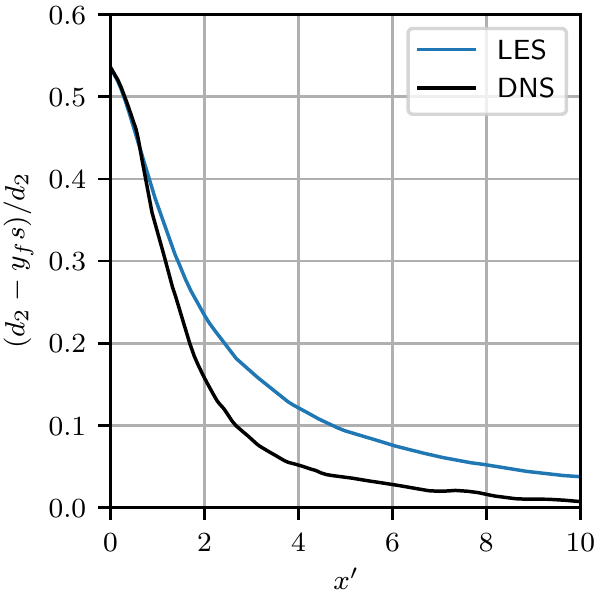}
	\caption{Convergence of the interface height towards $d_2$ in the benchmark simulation.}
	\label{fig:bench_interface_error}
\end{figure}

As discussed in the introduction, the depth of the water after the jump, $d_2$, can be computed a priori.
It is therefore possible to compute how the location of the interface approaches $d_2$ with increasing $x$.
The corresponding graph is shown in Figure~\ref{fig:bench_interface_error} along with the reference DNS data.
We note that the value at $x'= 0$ is fixed through the definition for $x_{toe}$, which explains why the agreement with the DNS is perfect.
The rate of growth of the water depth continues to be similar in both the LES and DNS up to $x' \approx 1$.
Further downstream the DNS values converge towards $d_2$ at a faster pace, and for the LES, full convergence is in fact not achieved in the limits of the computational domain.
The observed discrepancy is likely explained by the difference in the treatment of the outflow boundary.

\begin{figure}[htp!]
	\centering
	\includegraphics{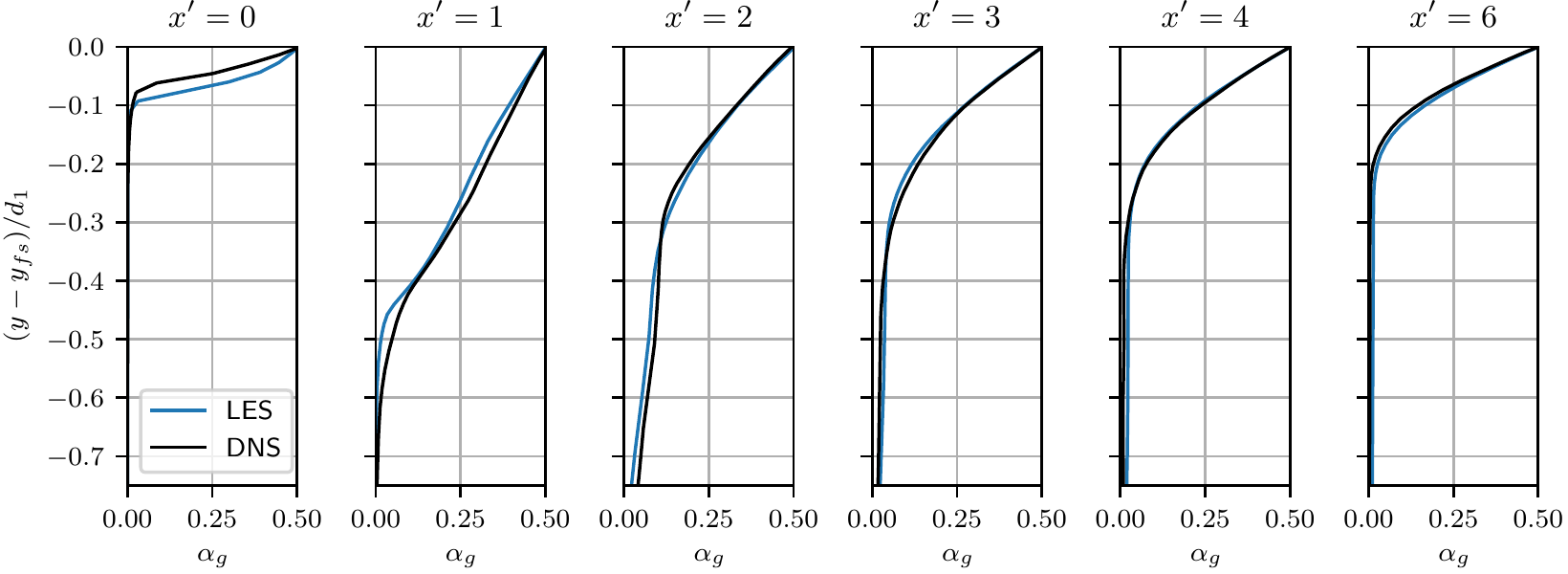}
	\caption{The profiles of $\mean{\alpha}$ in the benchmark simulation.}
	\label{fig:bench_alpha_1d}
\end{figure}

Figure~\ref{fig:bench_alpha_1d} shows the obtained profiles of $\mean{\alpha}$.
\rev{Agreement with the DNS is extremely good, with observable discrepancies only at $x'=0$ and $x'=1$.}
As discussed above, the most intense air entrainment occurs right downstream of the toe, so it is unsurprising that capturing the correct $\mean{\alpha}$ profile in this region is the most difficult.

\begin{figure}[htp!]
	\centering
	\includegraphics{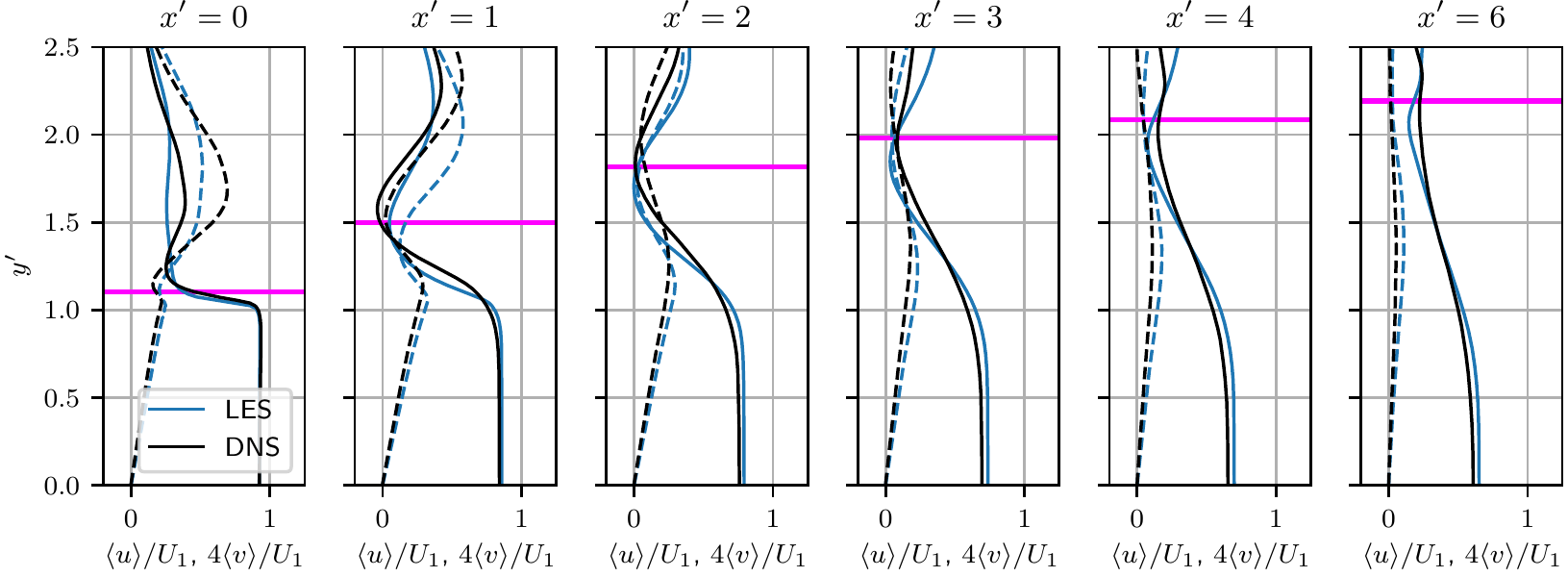}
	\caption{The profiles of $\mean{u}$ and $4\mean{v}$ obtained in the benchmark simulation. The magenta line shows the location of the interface.}
	\label{fig:bench_u_1d}
\end{figure}

The mean streamwise and vertical velocity profiles are shown in Figure~\ref{fig:bench_u_1d}.
The horizontal magenta lines show the positions of $\mean{\alpha_{0.5}}$.
Excellent agreement with the reference is obtained at all 6 streamwise positions.
Noticeable deviation is only observed in the values of vertical velocity of the air, which are not of particular interest and can be significantly affected by the boundary condition at the top of the domain.

\begin{figure}[htp!]
	\centering
	\includegraphics{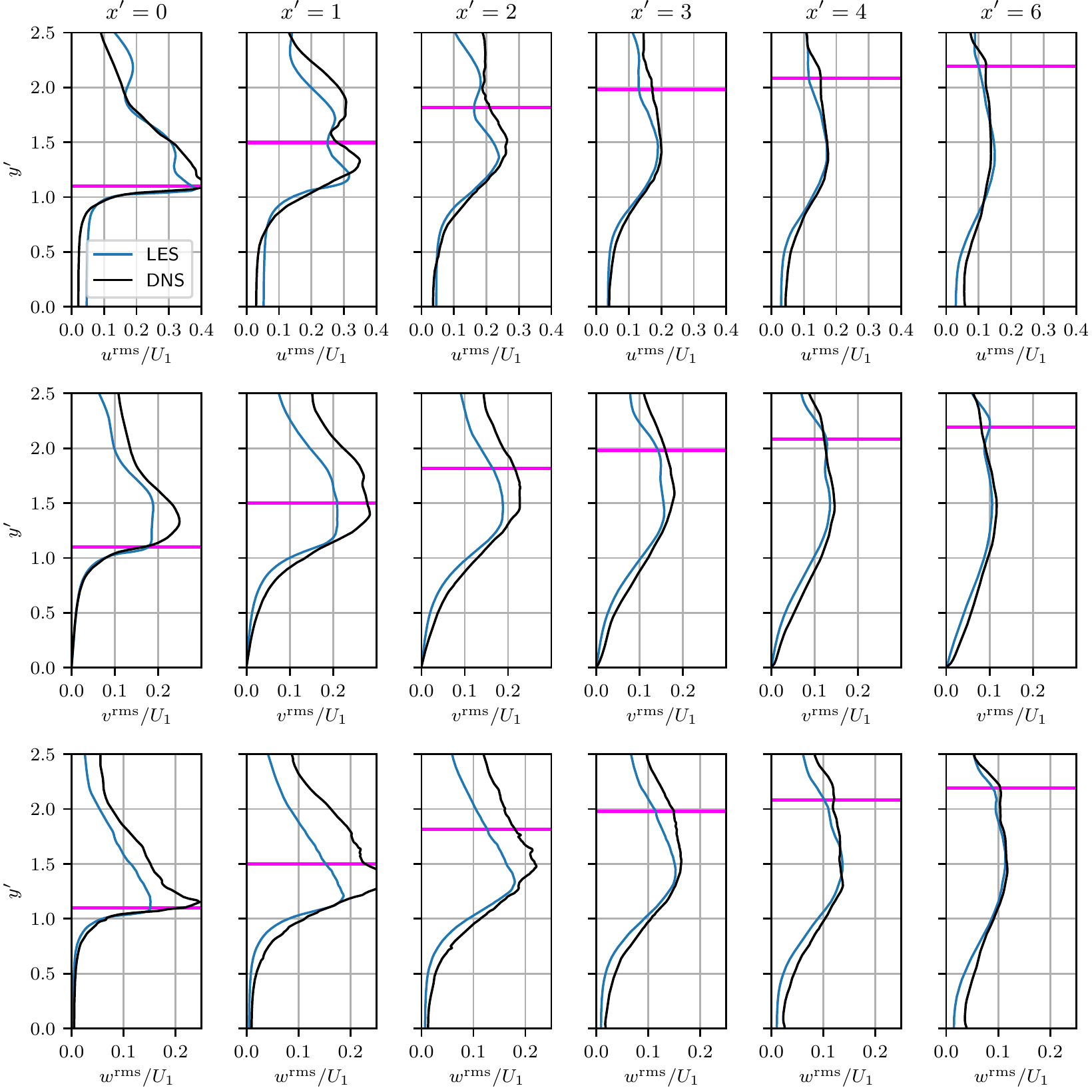}
	\caption{The profiles of $u^{rms}/U_1$, $v^{rms}/U_1$, and $w^{rms}/U_1$ obtained in the benchmark simulation. The magenta line shows the location of the interface.}
	\label{fig:bench_rms_1d}
\end{figure}

The profiles of the root-mean-square values of the three velocity components are shown in Figure~\ref{fig:bench_rms_1d}.
We note that inspection of the DNS data clearly shows that these second-order statistical moments are not completely converged, see Figure 8 in~\cite{Mortazavi2016a}.
In light of this, and the differences in the simulation setup, the obtained agreement is generally very good.
All three components are predicted with similar accuracy.
It is noteworthy that the disagreement with DNS is chiefly observed in the air and a short distance below the interface, whereas closer to the bottom the match is close to perfect.
In other words, the accuracy becomes worse in the presence of rapid interface topology changes.
This clearly demonstrates the importance of the interaction between the multiphase and turbulent structures.

\begin{figure}[htp!]
	\centering
	\includegraphics{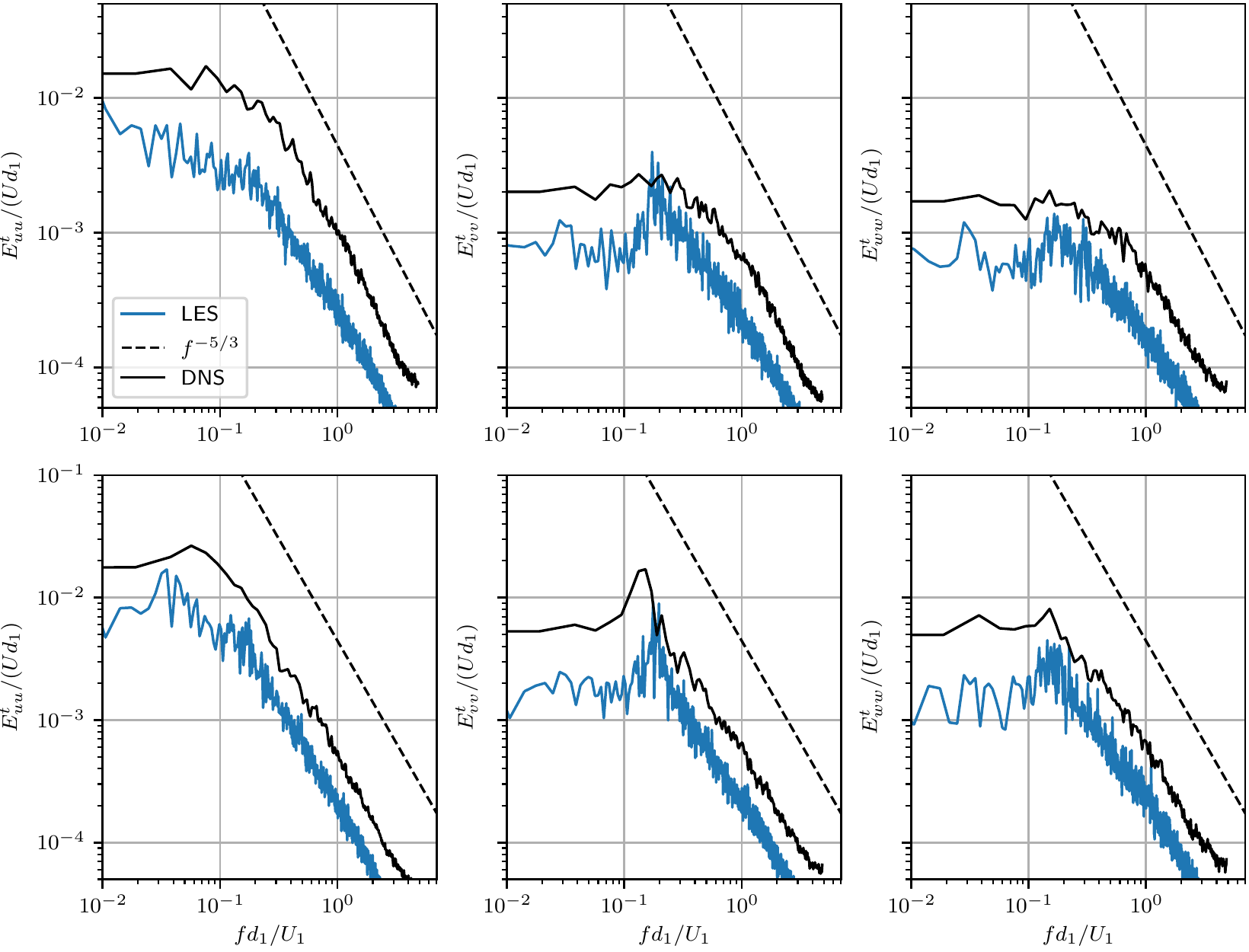}
	\caption{Temporal energy spectra of the three components of velocity at two selected $[x', y']$ positions: $[1.24, 1]$ (top), $[3.24, 1.1]$ (bottom).}
	\label{fig:bench_spectra}
\end{figure}

The analysis continues with the consideration of the temporal energy spectra of the velocity fluctuations.
These were computed at two $[x', y']$ positions: $[1.24, 1]$,  $[3.24, 1.1]$.
These are shown with red dots in top-left plot in Figure~\ref{fig:bench_alpha_2d}.
Note that the $x'$ values were essentially an outcome of the simulation, since it was not possible to know the value of $x_{toe}$ a priori.
Furthermore, the intention is to use the same $x$ and $y$ values in the whole simulation campaign, and the location of $x_{toe}$ varies slightly from simulation to simulation.
The values were therefore chosen in a conservative way to ensure that both locations are to the right of the toe.
The DNS data also provides temporal velocity spectra, including the following $[x', y']$ positions: $[0, 1]$,  $[2, 1.1]$.
Both the DNS and LES data are shown in Figure~\ref{fig:bench_spectra}.
The LES recovers the correct slope in the inertial range, which is in most cases close to the canonical $-5/3$-power spectrum.
Less energy is contained in the fluctuations in the case of the LES, but a this is likely to be a consequence of the signals being sampled from locations further from the toe.
Spectra for all three velocity components are predicted with comparable precision.

\begin{figure}[htp!]
	\centering
	\includegraphics{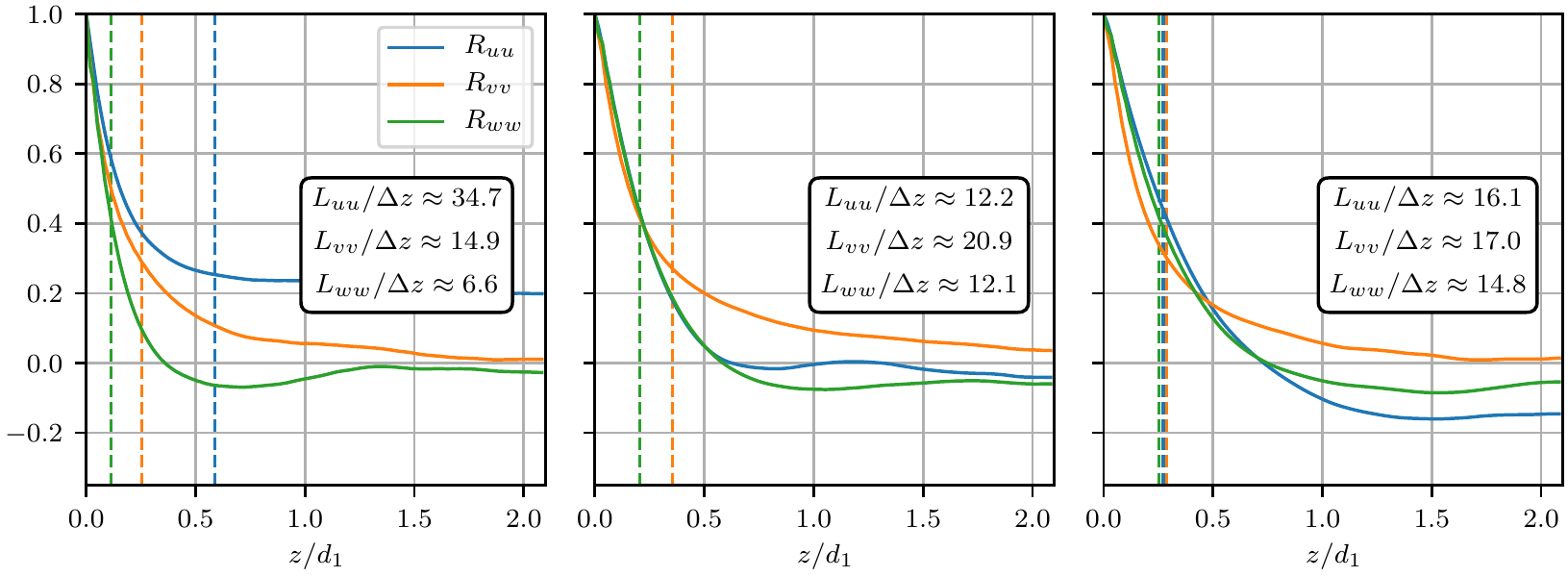}
	\caption{Spanwise two-point auto-correlations of velocity components computed at 3 selected $[x', y']$ locations: $[1.24, 1]$ (left),  $[3.24, 1.1]$ (middle), $[5.24, 1]$ (right).
	Vertical dashed lines show the integral length scale.}
	\label{fig:bench_corr}
\end{figure}

Next, the spanwise autocorrelation functions of the three velocity components, $R_{u_iu_i}$, are considered.
These are computed at the same two $[x', y']$ locations as the temporal spectra, plus an additional location further downstream: $[5.24, 1]$, see the black dot in Figure~\ref{fig:bench_alpha_2d}.
The result is shown in Figure~\ref{fig:bench_corr}.
Evidently, $R_{uu}$, does not decline to zero for two of the three considered locations.
This indicates that the spanwise dimension of the computational domain is somewhat insufficient, and prompted the use of a larger domain for the simulations on the $\Delta x3$ and $\Delta x4$ meshes.
The figure also presents the ratio of the integral length scales $L_{u_iu_i}$ and the cell size in the spanwise direction $\Delta z$.
The smallest scale to be discretized is $L_{ww}$, and at $[1.24, 1]$ it is only covered by $\approx 6.6$ cells.
By comparison, in~\cite{Davidson2009}, 8 cells is recommended for a \textit{coarse} LES.
This indicates that even with the $\Delta x1$ mesh some turbulent scales are resolved poorly.
Alternatively, the integral length scale may be a poor metric to relate grid resolution to for this particular flow.
In any case, further downstream $L_{u_iu_i}$ grow, meaning that the resolution with respect to them improves.

\begin{figure}[htp!]
	\centering
	\includegraphics{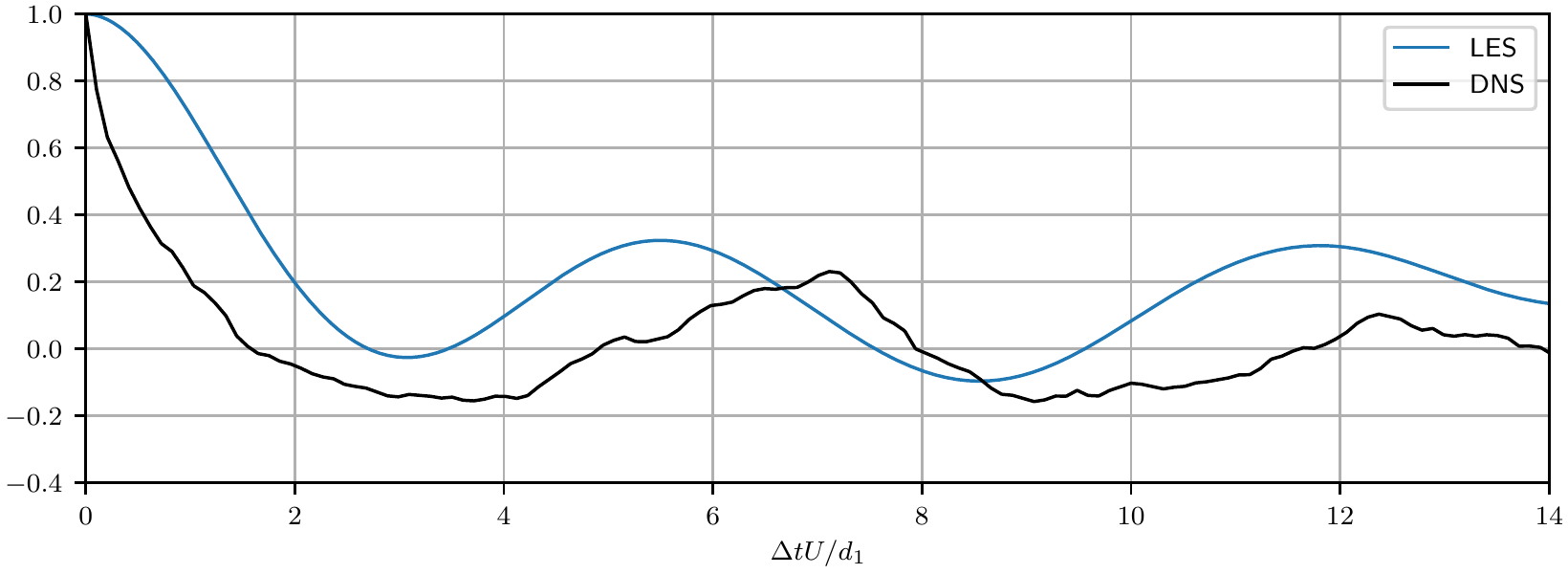}
	\caption{Autocorrelation function of the time-signal of the air volume passing through the box $x' \in [5.27, 6.09]$, $y' \in [0, 1.70]$.}
	\label{fig:bench_acf}
\end{figure}

Lastly, we analyse the air entrainment by considering the temporal variation of the volume of air passing through the box $x' \in [5.27, 6.09]$, $y' \in [0, 1.70]$.
The box is shown with red lines in the top-left plot in Figure~\ref{fig:bench_alpha_2d}.
Similar to the analysis made for the DNS~\cite{Mortazavi2016a}, we consider the autocorrelation function of the recorded signal.
The result is shown in Figure~\ref{fig:bench_acf}.
As expected, strong periodicity is revealed.
The DNS data appears somewhat unconverged, but the location of the first peak is relatively close to the LES.
The integral time-scales corresponding to the two curves are clearly different, but that is explained by the fact that the width of the box used for sampling the signal is larger in the LES.

The primary conclusion of this section is that OpenFOAM\textsuperscript{\textregistered} can be successfully used for scale-resolving simulations of the CHJ.
This can probably be extended to include other codes based on the same discretization and multiphase modelling frameworks.
In spite of the slight differences in the simulation setup, the observed overall agreement with the DNS data is good not only for first- and second-order statistical moments of the considered flow variables, but also for temporal turbulent spectra and air entrainment properties.
The largest deviations with DNS were observed directly downstream of the toe of the jump, which is physically the most complicated region to simulate.

\subsection{Influence of modelling parameters} \label{sec:parameters}

In this section, the effects of the grid resolution, amount of upwinding, and interface capturing method on the cost and accuracy of the results are considered.
The cost of the simulations is analysed first, and the associated metric, $N_h$ is defined as follows.
First, the simulation logs are used to compute the number of physical hours necessary to advance each simulation by 1 s.
Since the simulations on different grids were parallelised using different amounts of computational cores, the obtained timings are then multiplied by the corresponding amount of cores used.
This assumes linear scaling of computational effort with parallelisation, which is not exact, but provides a very good approximation in the range of core numbers used in the study.
Recall also that in the simulations using isoAdvector the time-step was adjusted to ensure the maximum Courant number is $< 0.5$, whereas $0.75$ was used in the MULES simulations.
To be able to account for the cost difference associated with the VoF algorithm as such, the cost metric for the MULES simulations was premultiplied by $0.75/0.5$.
Note that since a typical desktop computer has around 10 computational cores, and the full simulation needs to be run for about 10 s, $N_h$ also gives a rough estimate of how many hours it would take to perform a given simulation on a desktop machine.

The obtained values of $N_h$ are shown in Table~\ref{tab:cost}.
Each entry contains two numbers, corresponding to MULES and isoAdvector.
It is evident that the isoAdvector simulations are more expensive.
Depending on the other simulation parameters the ratio of $N_h$ varies within $\approx [1.17$, $1.55]$.
As a general trend, the isoAdvector becomes relatively more expensive with increased mesh resolution.
Numerical dissipation sometimes favourably affects the amount of iterations necessary to solve the pressure equation.
Here this effect is observed when the transition from 10\% to 25\% upwinding occurs, with the former always leading to a more expensive simulation.
However, for higher $u\%$, the effect of dissipaion on $N_h$  is neither particularly strong nor regular.
Considering the cost as a function of $\Delta x$, it is crucial to recall that the $\Delta x2$ simulations are performed on a thinner domain.
Since the computational effort does not scale linearly with the number of cells, this could not be directly accounted for in the metric.
Based on the data, on a desktop machine, it is possible to perform the $\Delta x4$ simulations in about 3 days, and the $\Delta x3$ in about 10.
For $\Delta x2$, the corresponding number is from 25 to 35 days depending on the simulation settings.
Taking into account the increased access of both academia and industry to HPC hardware, it can be said that the simulations on all three grids are relatively cheap, at least by LES standards.

\begin{table}[htp!]
	\caption{The simulation cost metric, $N_h$. For each $\Delta x$ and $u\%$ combination, two values are given, corresponding to MULES and isoAdvector, respectively.}
	\label{tab:cost}
	\begin{tabular}{lllll}
		& $u10\%$ & $u25\%$ & $u50\%$ & $u100\%$ \\ \hline
		$\Delta x2$ & $595$/$825$   & $553$/$855$ & $554$/$778$ & $591$/$801$ \\
		$\Delta x3$ & $284$/-       & $197$/$257$ & $199$/$264$ & $198$/$250$ \\
		$\Delta x4$ & $75$/-        & $53$/$62$   & $52$/$66$   & $50$/$64$   \\ \hline
	\end{tabular}

\end{table}

\begin{figure}[htp!]
	\centering
	\includegraphics{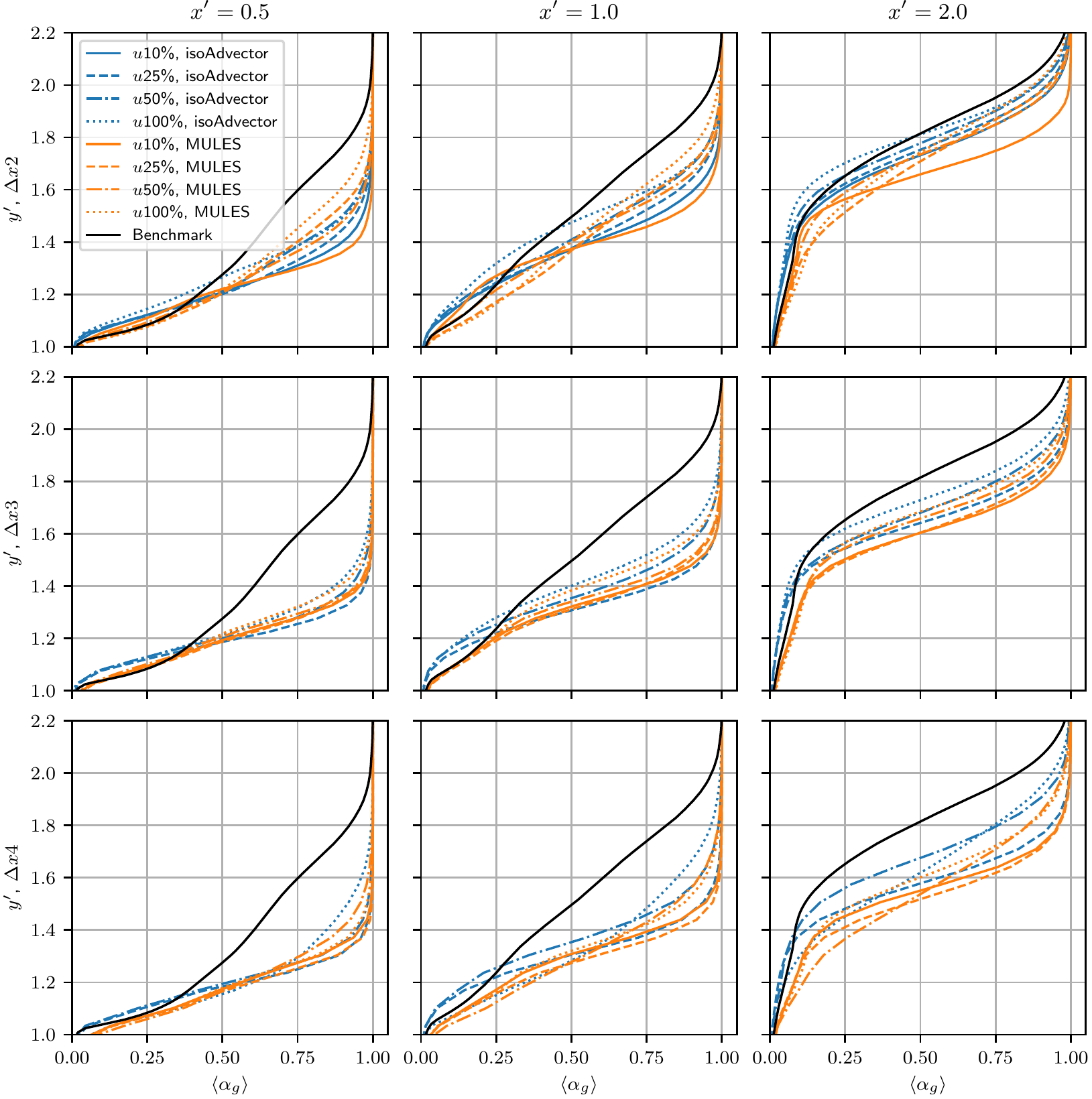}
	\caption{The profiles of $\mean{\alpha}$ obtained in the simulation campaign.}
	\label{fig:alpha_diss}
\end{figure}

Next, the computed profiles of $\mean{\alpha}$ are investigated, see Figure~\ref{fig:alpha_diss}.
The benchmark simulation revealed that the region of the flow that is most difficult to predict is directly downstream of the toe.
Therefore, here we focus on the following streamwise positions: $x'=0.5$, $1.0$, $2.0$.
The clear trend overarching all $x'$ and $\Delta x$ is that a higher amount of upwinding leads to better results.
\rev{For the majority of $\Delta x$ and streamwise positions, using isoAdvector
and $u100\%$ leads to the best predictive accuracy.}
The fact that using more dissipative schemes improves results is somewhat unexpected, because typically the recommendation for scale-resolving simulations is to keep dissipativity to a minimum.
However, it should be appreciated that in VoF any parasitic currents arising due to numerical errors of dispersive type propagate into errors in the advection of the interface.
It appears that avoiding these errors is more important than resolving steep velocity gradients.
As expected, the quality of the results degrades with the coarsening of the mesh.
The most precise result on $\Delta x2$ is quite close to the benchmark.
On the coarser grids, the accuracy is acceptable considering how cheap the corresponding simulations are.

\begin{figure}[htp!]
	\centering
	\includegraphics{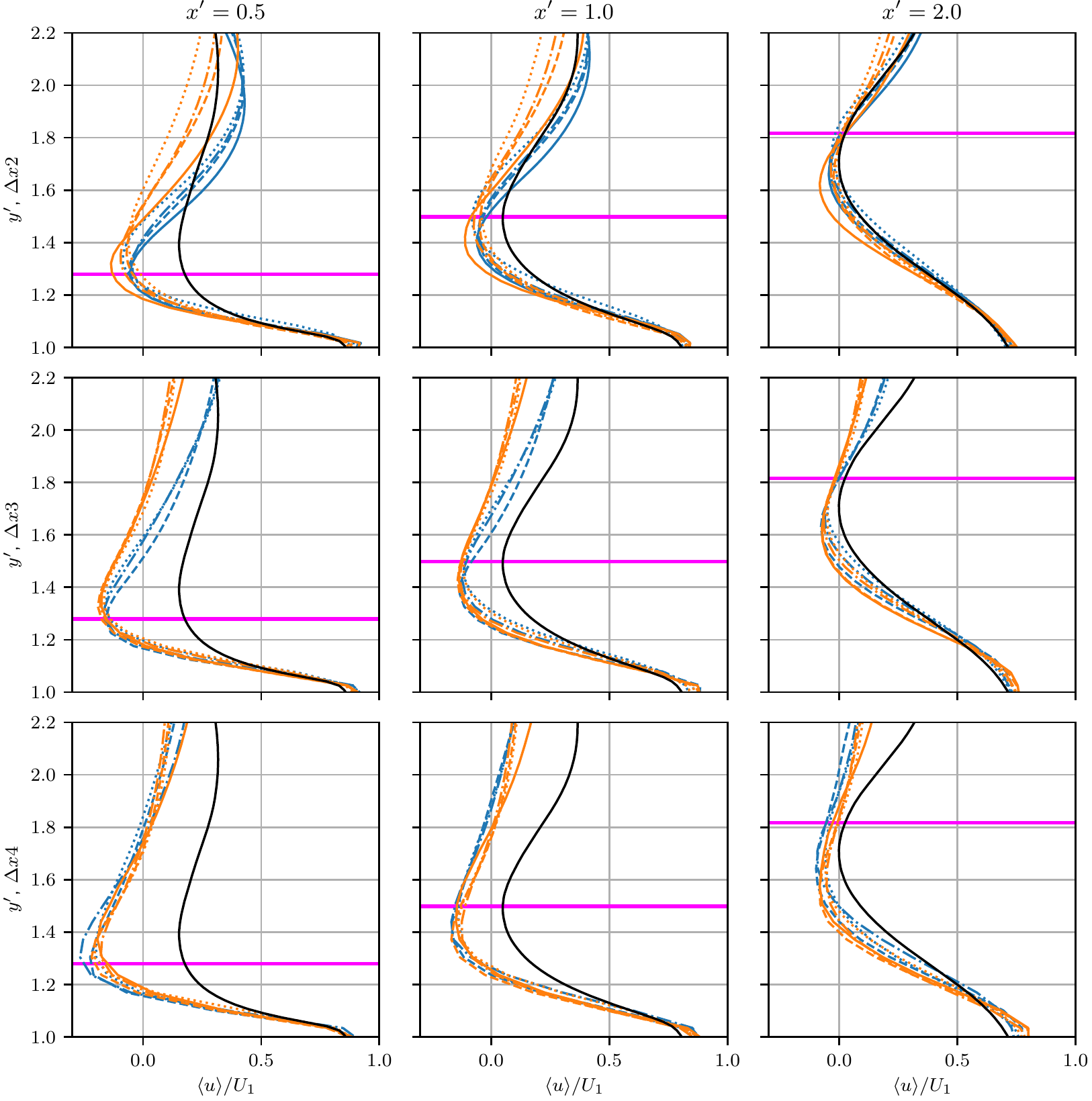}
	\caption{The profiles of $\mean{u}/U_1$ obtained in the simulation campaign. Line styles and colours as in Figure~\ref{fig:alpha_diss}.}
	\label{fig:u_diss}
\end{figure}

The predictions of the mean velocity are analysed next, see Figure~\ref{fig:u_diss}.
We focus on the streamwise component $\mean{u}$ only, since the level of accuracy of $\mean{v}$ is similar.
It is clear that compared to $\mean{\alpha}$, the results are more robust with respect to the amount of upwinding.
This is rather peculiar: The choice of interpolation scheme for $u$ has little effect on $\mean{u}$, but a stronger effect on a different quantity, $\mean{\alpha}$.
Nevertheless, the profiles obtained with higher $u\%$ are generally slightly more accurate, at least in the water phase.
Using isoAdvector leads to superior accuracy in the gas phase, whereas in the water phase no significant advantage over MULES is achieved.
The combination of $\Delta x2$, $u100\%$ and isoAdvector gives the best results, which are close to the benchmark.
At coarser resolutions accuracy deteriorates but not as strongly as for $\mean{\alpha}$.

\begin{figure}[htp!]
	\centering
	\includegraphics{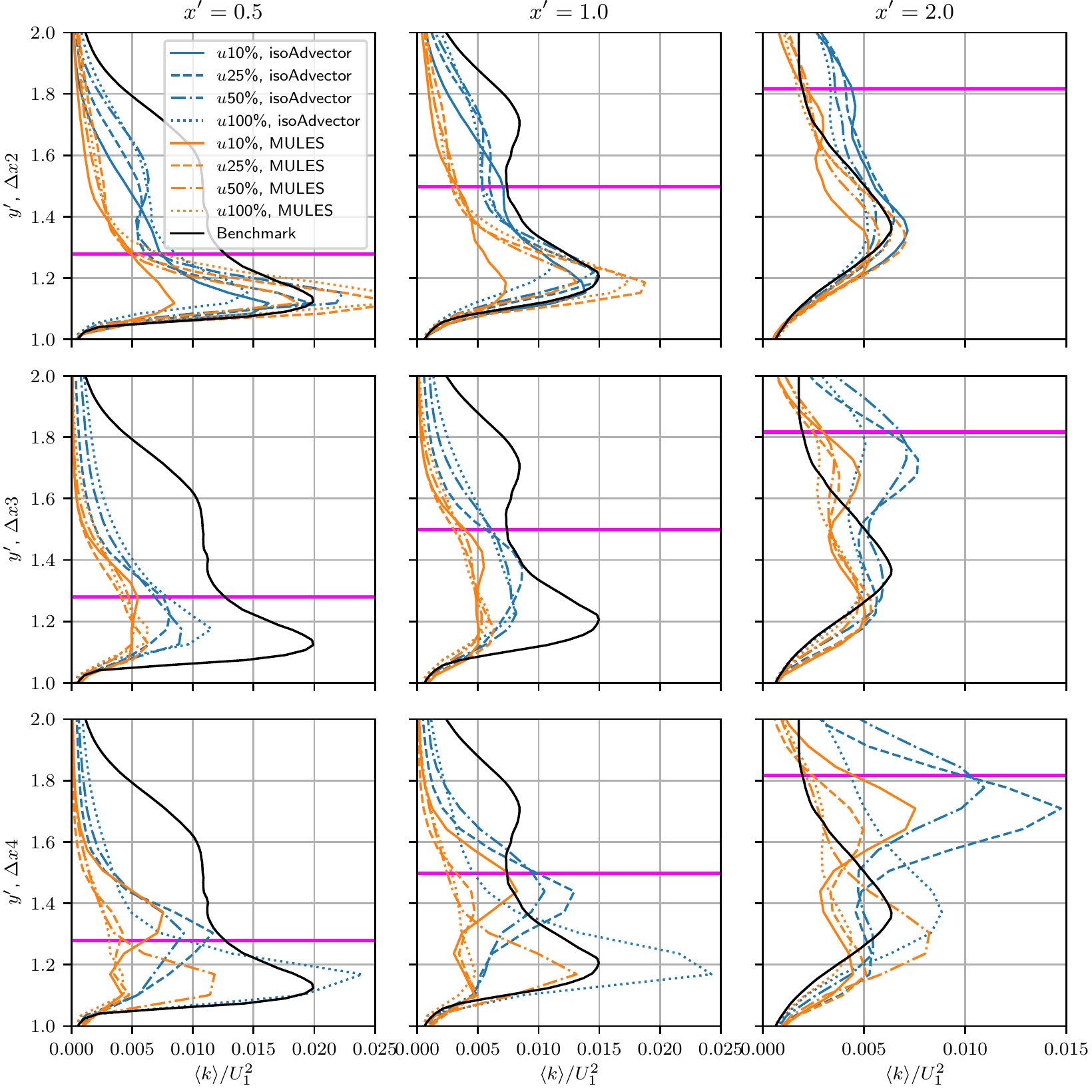}
	\caption{The profiles of $\mean{k}/U^2_1$ obtained in the simulation campaign.}
	\label{fig:k_diss}
\end{figure}

Figure~\ref{fig:k_diss} shows the obtained profiles of $\mean{k}$.
The observed error patterns are significantly less regular than in $\mean{u}$ and $\mean{\alpha}$.
Two factors contribute to this.
One is that $\mean{k}$ lumps together the errors in the variances of the three velocity components.
The other is that parasitic oscillations have a direct amplifying effect on $\mean{k}$.
Both of the above can lead to either error cancellation or amplification.
\rev{On the $\Delta x2$ grid, the best results are achieved with isoAdvector and 25/50\% upwinding.
In case of $u100\%$, the main peak in the detached shear layer is somewhat under-predicted, but the discrepancy is not very significant.}
An interesting observation is that at lower grid resolutions, a secondary peak in $\mean{k}$ is developed for $x'=1.0$ and $2.0$ right underneath the interface.
This unphysical peak is more pronounced when isoAdvector is used, and can even be observed on the $\Delta x2$ grid when this interface capturing technique is used.
It is present in all three components of the velocity variance, although for the streamwise component it is less pronounced.
The size of the peak grows with decreasing amount of upwinding, which confirms its numerical origin.
Even apart from this additional peak, the results for $\mean{k}$ on $\Delta x3$ and $\Delta x4$ are quite inaccurate, although the combination $\Delta x 4$, $u50\%$, MULES does reproduce the main features of the benchmark profiles fairly faithfully.

\begin{figure}[htp!]
	\centering
	\includegraphics{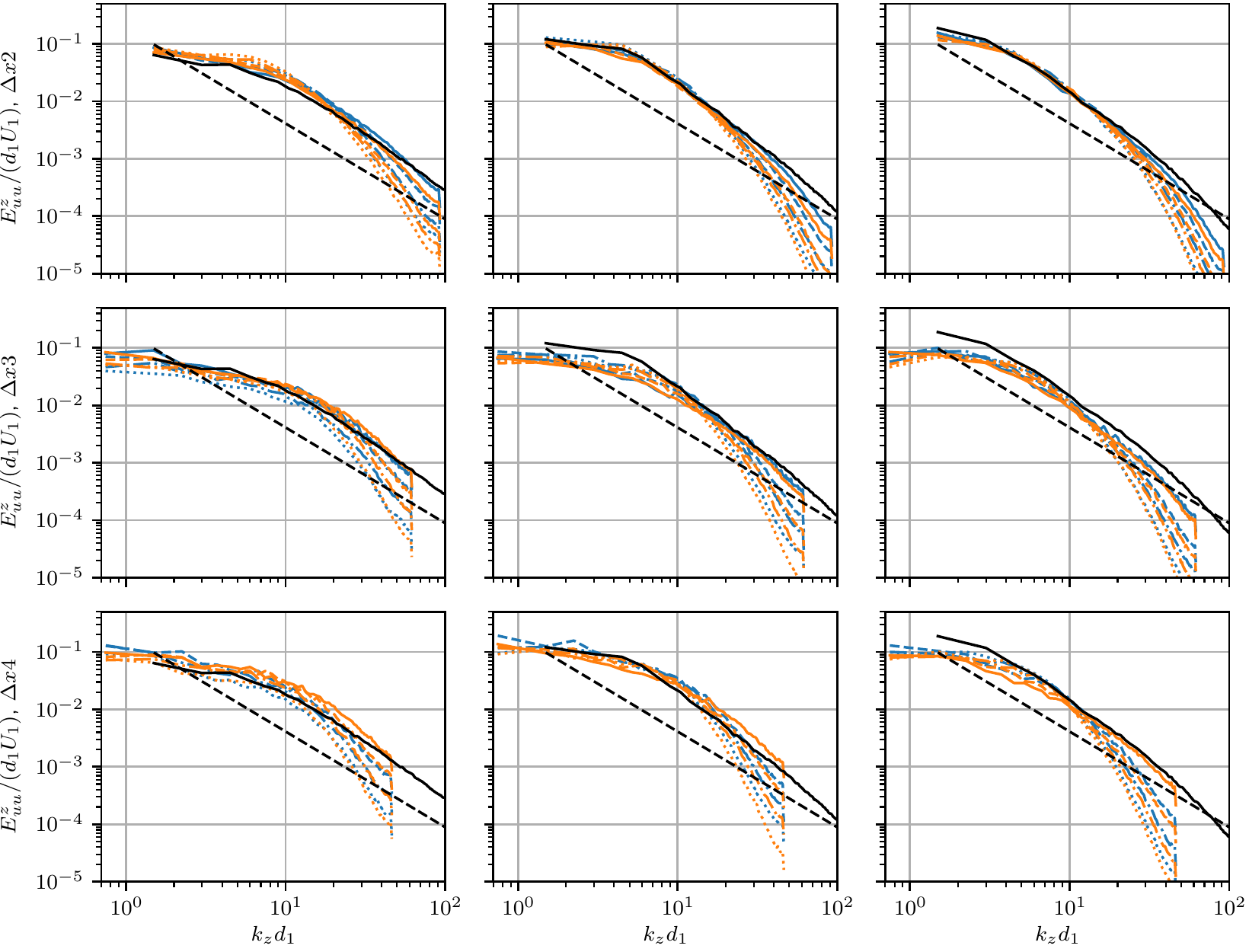}
	\caption{The spanwise velocity spectra obtained in the simulations at three selected locations: $[1.24, 1]$ (left), $[3.24, 1.1]$ (middle), $[5.24, 1.4]$ (right). Line styles and colours as in Figures~\ref{fig:alpha_diss} and \ref{fig:k_diss}.}
	\label{fig:specta_diss}
\end{figure}

The analysis of velocity predictions is now concluded with considering the spanwise energy spectra of the streamwise velocity, see Figure~\ref{fig:specta_diss}.
The spectra are computed at the same three $[x, y]$ locations as the spanwise autocorrelation functions for the benchmark simulations.
This entails that the respective $x'$ values are slightly different from simulation to simulation.
The reason for considering spanwise spectra instead of temporal is that, due to a larger amount of samples to average across, the spanwise spectra are much smoother, making it easier to distinguish the profiles from different simulations in the plots.
Unsurprisingly, increased upwinding leads to heavier dampening of the high-frequency fluctuations.
Due to the log-log scale being used, it is actually difficult to distinguish any effects of the VoF algorithm or $\Delta x$, besides for the fact that the frequency band of the spectrum is larger for denser meshes.
One could say that for small amounts of $u\%$ the spectrum is relatively well-predicted even at $\Delta x4$.
Therefore, one should exercise caution when making judgements regarding mesh resolution based on spectrum predictions.

\begin{figure}[htp!]
	\centering
	\includegraphics{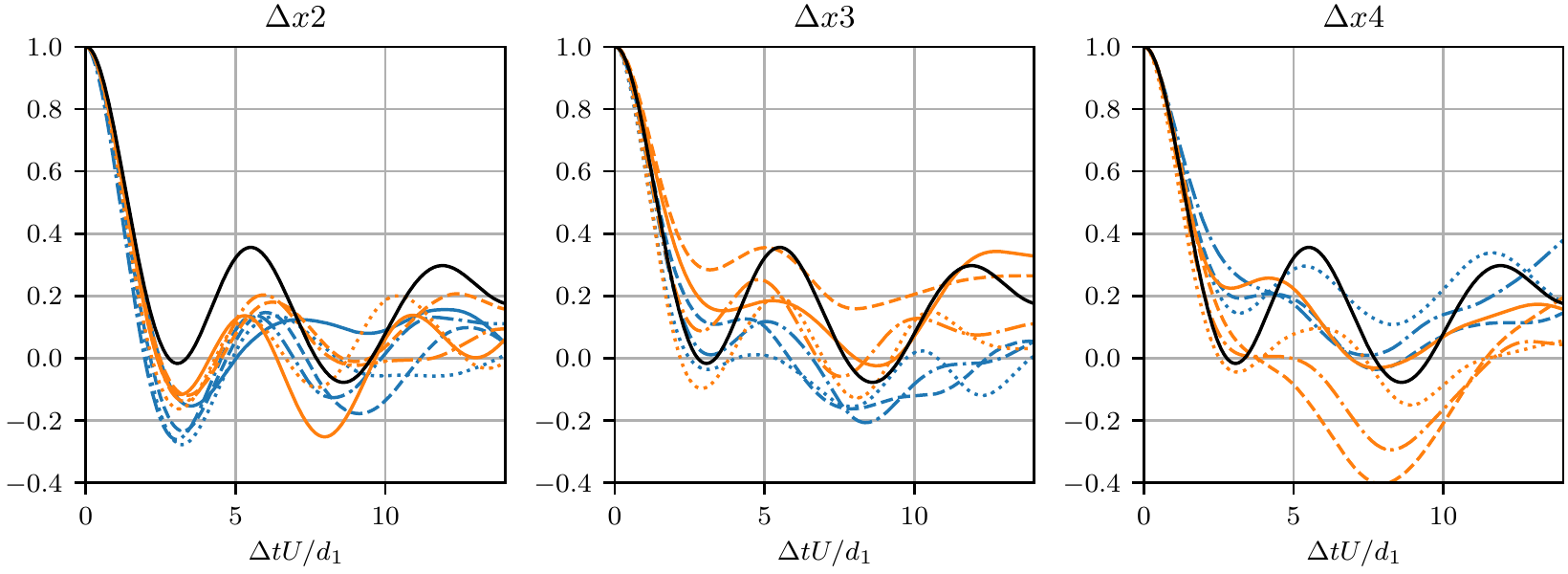}
	\caption{The obtained autocorrelation functions of the volume of leaked air.  Line styles and colours as in Figures~\ref{fig:alpha_diss} and \ref{fig:k_diss}.}
	\label{fig:acf_diss}
\end{figure}

Finally, the periodicity of air entrainment is analysed.
As for the benchmark simulation, the autocorrelation functions of the volume of air passing through a box located some distance downstream of the toe (see top-left plot in Figure~\ref{fig:bench_alpha_2d}) were computed.
The results are shown in Figure~\ref{fig:acf_diss}.
For $\Delta x2$ and $\Delta x3$, the location of the first peak is quite well predicted by all the simulations, whereas for $\Delta x4$ the accuracy deteriorates, in particular for some of the simulations using MULES.
Animations of the $\alpha = 0.5$ isosurface reveal that isoAdvector does a much better job at preserving the sharpness of the interface as the entrained bubbles travel downstream.
Therefore, if tracking the fate of the bubbles is important, using this VoF approach is recommended.
It should also be noted that while all the simulation predict similar entrainment frequencies, other statistical air entrainment properties do not agree equally well.
For example, the mean amount of air within the monitored box is highly affected by the choice of the VoF method, with MULES giving systematically higher values.

\section{Conclusions} \label{sec:conclusions}

This article presents results from an extensive simulation campaign studying the effects of different modelling parameters on the accuracy of LES of CHJ flow at $\text{Fr}_1 =2$.
The simulations were performed with a general-purpose finite-volume based CFD code, making the obtained results relevant for industry professionals and researches alike.

A benchmark simulation on a dense grid was conducted to test whether commonly employed VoF-based multiphase modelling methodologies are sufficiently accurate to capture the complicated physics of the flow.
Comparison with DNS data~\cite{Mortazavi2016a} has shown that the answer is positive, and good agreement with the reference has been found for the considered quantities of interest.
However, it was also revealed that numerical instabilities, discussed in Section~\ref{sec:instable}, constitute a significant problem.
It is virtually impossible to know a priori whether the chosen numerical setup would lead to a stable simulation, and a crash may occur sporadically after a significant part of the simulation time has already past.
Addressing the primary sources of instability (surface tension, density gradient term in~\eqref{eq:lesmom}) should therefore be a high priority \rev{for the development of VoF solvers in OpenFOAM\textsuperscript{\textregistered} and other codes based on similar algorithms}.

The rest of the simulation campaign focused on the effects of grid resolution, amount of upwinding, and VoF methodology.
One of the most interesting results is that the most dissipative scheme, $u100\%$, led to the best results for nearly all the considered quantities of interest.
Fortunately, dissipation also favours stability, which means that having both an accurate and stable numerical setup is possible.

Using the geometric VoF methodology, isoAdvector, was shown to lead to improved accuracy of the results and preservation of the interface sharpness.
This characteristic is particularly important if one of the simulation goals is tracking the fate of entrained bubbles.
However, the chance of instability is also increased by isoAdvector, and for some combinations of modelling parameters the simulations could not be run.
Unfortunately, this included the benchmark simulation.
The computational costs of isoAdvector simulations are also significantly larger than their MULES counterparts, see Table~\ref{tab:cost}.
For equivalent simulation settings, the maximum cost ratio was 1.5, however due to MULES being more stable it is possible to select a larger time step, which makes the difference even larger.

\rev{The combination of parameters that resulted in good predictions for all the quantities of interest is the $\Delta x2$ grid, the isoAdvector and the $u100\%$ scheme.
This combination of parameters is therefore recommended when compromising accuracy in favour of computational efficiency is not an option.
Using MULES instead is an alternative when guaranteed numerical stability and improved efficiency can motivate a modest reduction in the accuracy of the profiles and worsened resolution of individual bubbles.}
The $\Delta x3$ grid could be used to reduce costs significantly and still maintain a level of predictive accuracy that can be suitable for industrial simulations.
Using the $\Delta x4$ can only be recommended when the CHJ is a part of a larger flow configuration and is not of particular interest as such.

\section{Acknowledgements} \label{sec:acknowledgements}
This work was supported by grant number P38284-2 from the Swedish Energy Agency.
The simulations were performed on resources provided by Chalmers Centre for Computational Science and Engineering (C3SE) and UNINETT Sigma2---the National Infrastructure for High Performance Computing and Data Storage in Norway.
The authors are thankful to Milad Mortazavi for sharing the data files from the DNS simulation~\cite{Mortazavi2016a}.

\appendix
\bibliographystyle{plain}
\bibliography{library}


\printcredits

\end{document}